\documentstyle[epsfig,aps,preprint,tighten]{revtex}

\begin{document}
\preprint{nucl-th/9702015 (REVISED)}

\draft

\title{Polarization Observables in $\bbox{\phi}$ Meson Photoproduction
\\ and the Strangeness Content of the Proton}

\author{
Alexander I. Titov$^{a,b}$, Yongseok Oh$^{b,c}$, and Shin Nan Yang$^b$
}

\address{
$^a$Bogoliubov Laboratory of Theoretical Physics, JINR,
141980 Dubna, Moscow Region, Russia \\
$^b$Department of Physics, National Taiwan University,
Taipei, Taiwan 10617, Republic of China \\
$^c$Institut f\"ur Theoretische Physik, Physik Department,
Technische Universit\"at M\"unchen, D-85747 Garching, Germany \\
}
\date{\today}

\maketitle

\begin{abstract}
The contributions of direct knockout processes, in addition to the
diffractive production and one-pion-exchange processes associated with
the $\phi\rho\pi$ and $\phi\gamma\pi$ couplings, to the polarization
observables of the $\phi$ photoproduction from proton are calculated.
We make use of Pomeron-photon analogy and a relativistic harmonic
oscillator quark model.
We find that some of the double polarization observables are very
sensitive to a possible $s\bar s$ admixture in the proton.
It arises from the difference in the  spin structures of the three
different amplitudes.
This suggests that such measurements could be very useful to probe the
strangeness content in the proton.

\end{abstract}


Essentially all the constituent quark models which give good description
of the low energy properties of baryons treat nucleon as consisting of
only up and down quarks.
It naturally comes as a big surprise when some recent experiments and
theoretical analyses indicate a possible existence of a non-negligible
strange quark content in the proton.
For example, measurements of the nucleon spin structure functions
\cite{DIS} indicate that the amount of spin carried by the strange quark
pairs $s\bar s$ is comparable to that carried by the $u$ and $d$ quarks
and polarized opposite to the nucleon spin.
A similar conclusion has been drawn from the BNL elastic $\nu p$
scattering \cite{BNL}.
Analyses of the $\pi N$ sigma term \cite{sigmaterm} also suggest that
proton might contain an admixture of $20$\% strange quarks.
Recently, experiments on annihilation reactions $\bar p p \to \phi X$
($X= \pi^0, \eta, \rho^0, \omega, \gamma$) at rest \cite{pp} show a
strong violation of the OZI rule \cite{OZI}.
Again, the presence of an $s\bar s$ component in the nucleon wave
function would enable the direct coupling to the $\phi$ meson in
the annihilation channel without violating the OZI rule \cite{EKKS,GFYY}.
However, it has also been argued that such experimental results could
be understood with little or no strangeness in the nucleon
\cite{others1,others2}.
Resolution of this question will provide us with important insight to
improve current hadron models.
It would be hence important and interesting to look into other possible
clear signal \cite{newexp} that might be related directly with the
strangeness content of nucleon.

\begin{figure}
\centerline{\epsfig{file=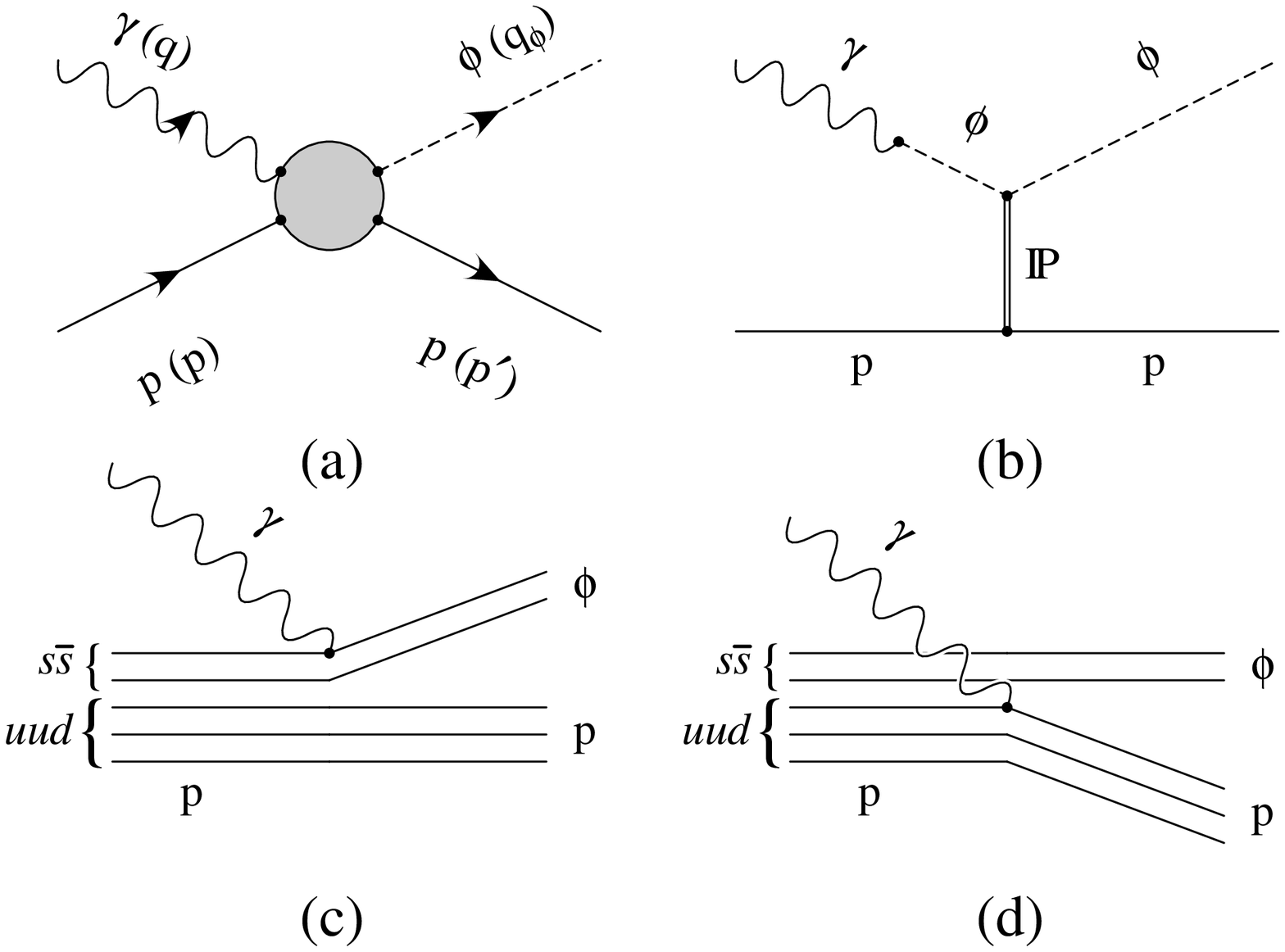, width=0.7\columnwidth}}
\caption{(a) $\phi$ meson photo- and lepto-production from proton, where
the photon would be virtual for the latter case.
(b) diffractive $\phi$ production of vector-meson dominance model.
(c,d) direct knockout mechanism to the $\phi$ production.}
\label{fig:dg}
\end{figure}

In the lepto- and photo-production of $\phi$ meson from proton of
Fig. \ref{fig:dg}(a), in addition to the diffractive
production of vector-meson dominance model (VDM) depicted in
Fig. \ref{fig:dg}(b), an $s \bar s$ admixture in the nucleon wave function,
if exists, can contribute to the process through the direct ``knockout''
mechanism of Fig. \ref{fig:dg}(c,d).
We use $p = (E_p, {\bf p})$, $p' = (E_p', {\bf p}')$,
$q = (\nu, {\bf q})$, and $q_\phi = (E_\phi, {\bf q}_\phi)$
to denote the four-momenta of the target and recoil proton, photon beam,
and produced $\phi$ meson in the laboratory frame, respectively.
The corresponding variables in c.m. system will be distinguished by an
asterisk as $p^*$, $p^{*\prime}$, $q^*$, and $q_\phi^*$.
In Ref. \cite{HKW}, this idea was applied to $\phi$ electroproduction
with a non-relativistic quark model, and the calculation was improved in
Ref. \cite{TYO1} by including relativistic Lorentz-contraction effects.
With the form factors and overlap integrals evaluated with a relativistic
harmonic oscillator quark model \cite{RHOM}, it was found that with
less than 5\% admixture of strange sea quarks in the proton, the cross
section of the direct knockout mechanism is comparable to that of VDM for
electroproduction at moderately large electron four-momentum transfer.
However, it is not easy to disentangle the two mechanisms from the cross
section measurement because their respective contributions  have similar
dependence on the momentum transfer \cite{TYO1}.

In this Letter, we propose that measurements of polarization observables
of $\phi$ meson photoproduction could offer a possible clean signature
of the strangeness content of nucleon.
Based on the model of Ref. \cite{TYO1}, we will show that some of the
double polarization observables are very sensitive to the hidden
strangeness content of proton, even with less than 1\% admixture of
$s \bar s$ component in the proton wave function.
This is because the contributions from the direct knockout and the
diffractive processes to these observables have very different spin
(helicity) dependence.

The diffractive $\phi$ photoproduction mechanism of VDM assumes
that the incoming photon mixes into the $\phi$ meson and then
scatters diffractively with proton through the exchange of
a Pomeron \cite{VDM}.
Experimental observations for the vector meson production, small-$|t|$
elastic scattering, and diffractive dissociation indicate that the
Pomeron behaves rather like an isoscalar photon-like particle.
A microscopic description of the vector-meson photoproduction at
high energy based on the Pomeron-photon analogy was proposed
by Donnachie and Landshoff \cite{DL}, and the Pomeron could be
described successfully in terms of non-perturbative two gluon
exchange \cite{DL,Cudel,Gol93,Laget}.
With the use of the spin structure of the diffractive $\phi$
photoproduction mechanism of these models, the invariant amplitude of
the diffractive production reads
\begin{eqnarray}
  T_{fi}^{\rm VDM} &=& i\, T_{0}\,
  {\bar u}(p') \gamma_\alpha u(p) \,
  \varepsilon_\mu^*(\phi) \Gamma^{\alpha,\mu\nu} \varepsilon_\nu(\gamma),
  \nonumber \\
  \Gamma^{\alpha,\mu\nu} &=&
  (q+q_\phi)^\alpha\, g^{\mu\nu} - q^{\mu}\, g^{\alpha\nu}
  - q^{\nu}_{\phi}\, g^{\alpha\mu},
\label{eq:VDM}
\end{eqnarray}
where $T_0$ includes the dynamics of Pomeron-hadron interaction and
is taken to be real~\cite{Pomeron}; $\varepsilon_\mu(\phi)$ and
$\varepsilon_\mu(\gamma)$ are the polarization vectors of $\phi$ meson and
photon, respectively, and $u(p)$ is the proton Dirac spinor.
The spin structure of the amplitude $T_{fi}^{\rm VDM}$ is the same
as that of a one-photon-exchange amplitude \cite{BD}.
$T_0$ is determined by fitting the data in the form of
$(d\sigma/dt)_{\rm VDM} = \sigma_\gamma(W) b_\phi
\exp( -b_\phi | t - t_{\rm max} | )$ as in Ref. \cite{Bonn} with
$b_\phi = 4.01$ GeV$^{-2}$ and $\sigma_\gamma(W) = 0.2$ $\mu$b around
$W = 2$ GeV, where $t = (p-p')^2$ and $W^2=(p+q)^2$.

\begin{figure}
\centerline{\epsfig{file=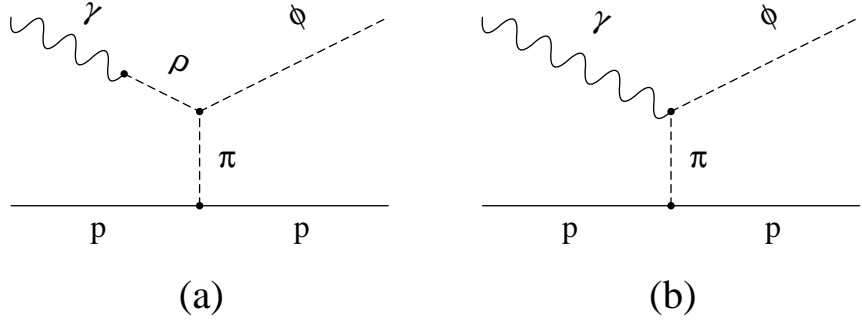, width=0.7\columnwidth}}
\caption{One-pion-exchange model (OPE) of $\phi$ photoproduction.}
\label{fig:ope}
\end{figure}

At relatively low photon energy, one-pion-exchange (OPE) diagrams of
direct $\phi$ photoproduction as shown in Fig. \ref{fig:ope} also
contribute \cite{VDM}.
It may be interpreted as a correction to the VDM process \cite{JMLS}.
It is straightforward to write down the amplitudes of the OPE diagrams
of Fig. \ref{fig:ope}.
They have an identical form and we may write their sum as
\begin{eqnarray}
  T_{fi}^{\rm OPE} &=& \frac{i}{t-m_\pi^2} g_{NN\pi} \tilde g_{\phi\gamma\pi}
  \bar u(p') \gamma_5 u(p)
\nonumber \\ && \mbox{} \times
  \epsilon^{\mu\nu\alpha\beta}
  q_{\phi,\mu} q_\alpha \varepsilon^*_\nu (\phi) \varepsilon_\beta (\gamma),
\label{amp-ope}
\end{eqnarray}
where $\tilde g_{\phi\gamma\pi}$ is the effective coupling constant
determined from the branching ratio of $\phi \to \gamma \pi$ and $m_\pi$
is the pion mass.
We also include the Benecke-D\"urr form factors for each vertex as given
in Ref. \cite{JMLS}.

The main ingredient of the knockout photoproduction mechanism is the
assumption that the constituent quark wave function of proton, in addition
to the usual 3-quark ($uud$) component, contains a configuration with
explicit $s \bar s$-pair.
A simple realization of this picture is the following wave function in Fock
space \cite{HKW}
\begin{eqnarray}
  |p \rangle = A | [uud]^{1/2} \rangle
  + B \Bigl\{ &a_0& | \bbox{[} [ uud]^{1/2}
  \otimes [s\bar s]^0 \bbox{]}^{1/2} \rangle
\nonumber \\
  \mbox{} + &a_1& | \bbox{[} [ uud]^{1/2}
  \otimes [s\bar s]^1 \bbox{]}^{1/2} \rangle \Bigr\},
\label{eq:wf}
\end{eqnarray}
where $B^2$ is the strangeness admixture of the proton and
$(a_0^2, a_1^2)$ are the fraction of the $s\bar s$ pair with spin 0 and 1,
respectively.
The superscripts represent the spin of each  cluster and the circle-cross
represents the vector addition of spins of $uud$ and $s\bar s$ clusters
and their relative orbital angular momentum ($\ell=1$).
Details on the wave functions in the relativistic harmonic oscillator
model \cite{RHOM} and electromagnetic current associated with the $\phi$
photo- and electro-production can be found in Refs.~\cite{HKW,TYO1}.

We classify the knockout mechanism into $s \bar s$- and $uud$-knockout
depending on the struck quark by the photon.
The knockout amplitudes are most easily evaluated in the laboratory frame
with the use of wave function (\ref{eq:wf}) as presented in Ref. \cite{TYO1}.
After transformation into c.m. frame, we have
\begin{eqnarray}
  T_{m_f,m_\phi;m_i,\lambda_\gamma}^{s \bar s} &=&
  i\, T_0^{s \bar s} {\cal S}^{s \bar s}_{m_f,m_\phi;m_i,\lambda_\gamma},
\nonumber \\
  T_{m_f,m_\phi;m_i,\lambda_\gamma}^{uud} &=&
  i\, T_0^{uud} {\cal S}^{uud}_{m_f,m_\phi;m_i,\lambda_\gamma},
\label{t_0}
\end{eqnarray}
where $m_{f,i}$, $m_{\phi}$, and $\lambda_\gamma$ are the spin projections
in $z$-direction of the final (initial) proton, $\phi$ meson, and photon
helicity, respectively, and
\begin{eqnarray}
  T_0^{s \bar s} &\propto&
  A^* B a_0 \, F_{s \bar s} (q_{s \bar s}) F_{uud} (0) V_{s \bar s} (p'),
\nonumber \\
  T_0^{uud} &\propto& - A^* B a_1 \, F_{s \bar s} (0) F_{uud} (q_{uud})
  V_{uud}(q_\phi).
\label{T:KO}
\end{eqnarray}
$F_{\alpha}(q_\alpha)$'s ($\alpha = s \bar s, uud$) are the Fourier
transforms of the overlap integrals between spatial wave functions of
the struck cluster $\alpha$ in entrance and exit channels with effective
momentum $q_\alpha$, which reduces to ${\bf q}^2$ in the non-relativistic
limit.
$V_\alpha$ is the momentum distribution function of cluster $\alpha$.
The explicit expressions of Eq. (\ref{T:KO}) are given in Ref.~\cite{TYO1}.
The spin structure functions are
\begin{eqnarray}
  {\cal S}^{s \bar s}_{m_f,m_\phi;m_i,\lambda_\gamma} &=&
  \sqrt3 \, \lambda_\gamma \sum_\varrho \,
  \langle {\textstyle\frac12}\, m_f \, 1\, \varrho\, | \,
  {\textstyle \frac12} \, m_i \rangle \,
\nonumber \\ && \mbox{} \times
  \bbox{\varepsilon}^* (m_\phi) \cdot \bbox{\varepsilon} (\lambda_\gamma)\,
  \xi^{s \bar s}_\varrho ,
\nonumber \\
  {\cal S}^{uud}_{m_f,m_\phi;m_i,\lambda_\gamma} &=&
  \sqrt3 \, \sum_{j_c,m_c,\varrho}
  \langle {\textstyle \frac12}\, m_f-\lambda_\gamma\, 1\, \varrho |
  j_c\, m_c \rangle\,
\nonumber \\ && \mbox{} \times
  \langle j_c\, m_c\, 1\, m_\phi | {\textstyle \frac12}\, m_i \rangle \,
  \xi_\varrho^{uud},
\end{eqnarray}
where
\begin{eqnarray}
  && \xi^{s \bar s}_{\pm 1} = \pm
  \frac{1}{\sqrt2} \sin\theta_{p'},
  \quad \xi^{uud}_{\pm 1} = \mp
  \frac{1}{\sqrt2} \sin\theta_{q_\phi},
\nonumber \\
  && \xi^{s \bar s}_0 = \cos\theta_{p'},
  \quad \xi^{uud}_0 = \cos\theta_{q_\phi},
\end{eqnarray}
with the production angle $\theta_k$ in the laboratory frame.

The corresponding amplitudes in helicity basis can be obtained with
the relation \cite{JW,POL},
\begin{eqnarray}
  H_{\lambda_f,\lambda_\phi;\lambda_i,\lambda_\gamma} &=&
  (-1)^{1-\lambda_i-\lambda_f} \sum_{m_i,m_f,m_\phi}
  d^{1/2}_{m_i,-\lambda_i} (0)\,
\nonumber \\ && \hskip -1cm \mbox{} \times
  d^{1/2}_{m_f,-\lambda_f} (\theta)\,
  d^1_{m_\phi,\lambda_\phi} (\theta) \,
  T_{m_f,m_\phi;m_i,\lambda_\gamma},
\label{t_b}
\end{eqnarray}
where $\theta$ is the c.m. scattering angle and $\lambda_{i,f,\phi}$ is
the helicity of the target (recoil) proton and $\phi$, respectively.
Note that the knock-out amplitudes of Eq.~(\ref{t_0}) are purely imaginary,
which means that the incoming photon is absorbed by the 5-quark component
of the target proton, whereas the OPE contribution of Eq.~(\ref{amp-ope})
is purely real.
The total photoproduction helicity amplitude $H$ is then the sum,
$H = H^{\rm VDM} + H^{\rm OPE} +  H^{s \bar s} + H^{uud}$.
Close inspection of the amplitudes reveals that each amplitude exhibits
different helicity structure. In particular, at $|t| \sim |t|_{\rm min}$
(i.e., $\theta \to 0$), we have
$H^{\alpha} \propto \Lambda_\alpha \delta_{\lambda_f,\lambda_i}
\delta_{\lambda_\phi,\lambda_\gamma}$,
where $\Lambda_{\rm VDM} = i$,
$\Lambda_{\rm OPE} = 2\lambda_i \lambda_\gamma$, and
$\Lambda_{s\bar s} = 2i\lambda_i \lambda_\gamma$,
while $H^{uud}$ is suppressed.
This gives rise to a strong interference between $H^{\rm VDM}$ and
$H^{s \bar s}$ at forward scattering region, while $H^{\rm VDM}$ and
$H^{uud}$ interfere strongly at large $\theta$.
The OPE amplitude $H^{\rm OPE}$ contributes incoherently to the unpolarized
cross section and the double polarization observables of our interest, while
the interference of $H^{\rm VDM}$ and the knockout amplitudes gives very
distinct contributions to the polarization observables.

\begin{figure}
\centerline{\epsfig{file=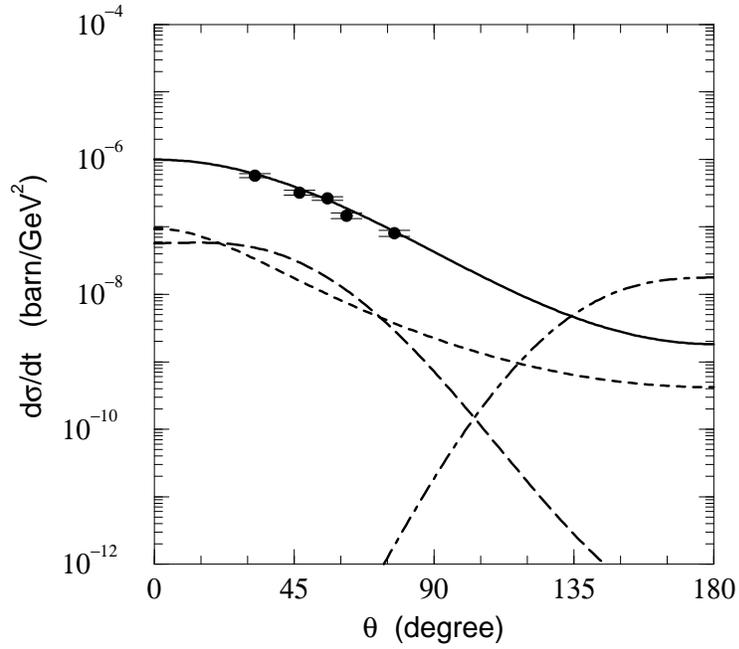, width=0.70\columnwidth}}
\caption{
The unpolarized photoproduction cross section
$d\sigma/dt (\theta)$ at $W = 2.155$ GeV
($E_\gamma^{\rm lab} = 2.0$ GeV).
The solid, dotted, dashed, and dot-dashed lines give the cross section
of VDM, OPE, $s \bar s$-knockout, and $uud$-knockout with strangeness
admixture $B^2 = 1$\% and $|a_0| = |a_1| = 1 / \protect\sqrt{2}$.
The experimental data are from Ref. \protect\cite{Bonn}.}
\label{fig:unpol}
\end{figure}

Figure \ref{fig:unpol} gives, together with the data of Ref. \cite{Bonn},
the unpolarized $\phi$ photoproduction cross sections $d\sigma/dt$ of
various mechanisms at $W = 2.155$ GeV as functions of $\theta$, assuming
that the strangeness probability $B^2 = 1$\% and $|a_0| = |a_1| = 1 / \sqrt2$.
One sees that the main contribution comes from VDM mechanism except at
large $\theta$, where $uud$-knockout dominates.
Excluding this extreme limit, where the validity of VDM is suspected
and the contributions from excited intermediate states are expected,
VDM process dominates  and the knockout contribution is small,
so the unpolarized cross section is not sensitive to the small
strangeness content of proton.
The OPE contribution is also small compared with the VDM, even
though it is comparable to or larger than that of the $s\bar s$-knockout.
This validates our choice of VDM parameters.

\begin{figure}
\centerline{\epsfig{file=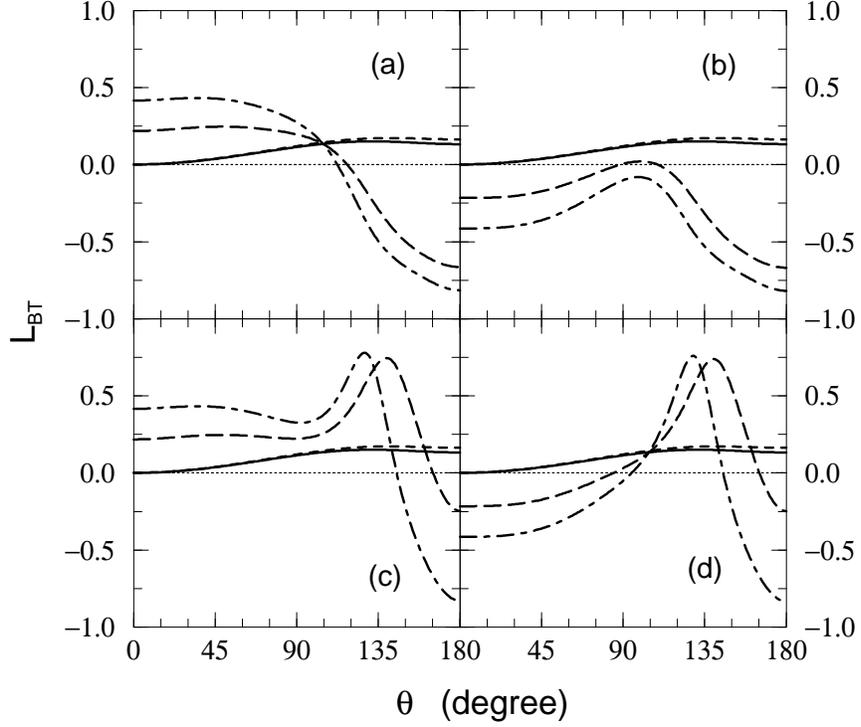, width=0.70\columnwidth}}
\caption{
The longitudinal asymmetry ${\cal L}_{BT} (\theta)$ at
$W = 2.155$ GeV with $B^2=0$\%, i.e., the VDM and OPE (solid lines),
0.25\% (dashed lines), and 1\% (dot-dashed lines) assuming
that $|a_0| = |a_1| = 1 / \protect\sqrt{2}$.
The dotted line, which nearly overlaps the solid line, is the prediction
of pure VDM.
The phases of $a_0$ and $a_1$ for (a), (b), (c), and (d)
are $(+,+)$, $(-,+)$, $(+,-)$, and $(-,-)$, respectively.}
\label{fig:bt}
\end{figure}

{}From the helicity amplitudes we can obtain various single, double,
or triple polarization asymmetries \cite{POL}.
It turns out that the single polarization asymmetries are not
sensitive to the strange quark admixture of proton while some of
double polarization asymmetries do.
We focus only on beam--target asymmetry ${\cal L}_{\rm BT}$ here
and the others will be reported elsewhere.
For photon and target proton  polarized along
$\pm \hat{\bf z}$ and $-\hat{\bf z}$, respectively, where
$\hat{\bf z} = \hat{\bf q}^*$, i.e., a longitudinal asymmetry, we have
\begin{eqnarray}
  {\cal L}_{\rm BT}
  \equiv \frac{|H_{u,u;+,+}|^2 - |H_{u,u;+,-}|^2}
              {|H_{u,u;+,+}|^2 + |H_{u,u;+,-}|^2},
\end{eqnarray}
where the subscripts $u$ and $\pm$ refer to unpolarized and
$\lambda_{i,\gamma} = \pm\frac 12$ case, respectively.
Shown in Fig. \ref{fig:bt} are our results for ${\cal L}_{\rm BT}$ at
$W = 2.155$ GeV.
Since there is an uncertainty in the phase between VDM and knockout
amplitudes, we give the results for four different choices for the
signs of $a_0$ and $a_1$ while keeping $|a_0|=|a_1|$.
We see that this polarization observable is useful to determine the
phase and magnitude of $a_0$.
This conclusion holds also for the longitudinal beam--recoil asymmetry
${\cal L}_{BR}$.
Unlike the unpolarized cross section, ${\cal L}_{\rm BT}$ strongly depends
on the strangeness content of proton.
Even with $B^2=0.25$\%, the difference with the (VDM $+$ OPE) prediction
is significant.
In addition, at small $\theta$ the results are nearly independent of the
phase of $a_1$, while they are nearly independent of the phase of $a_0$
at large $\theta$.
This is because the $uud$- ($s\bar s$-) knockout process is suppressed as
compared with the $s \bar s$- ($uud$-) knockout at small (large) $\theta$
region.
It is found that the effect of OPE is small for this asymmetry.

In summary, we find that, with the use of a relativistic harmonic
oscillator quark model, the direct knockout processes give a very
distinct contribution to some of the double polarization observables
in $\phi$ photoproduction as compared with those of diffractive production
and one-pion-exchange processes.
It indicates that measurements of double polarization observables
would be very useful in probing the strangeness content of the proton.
We also find that the contribution of the knockout mechanism is
suppressed with increasing initial photon energy because of the
strong suppression of form factors in the knockout amplitudes.
Therefore, the optimal range of the initial photon energy to measure
the $s \bar s$ component of proton is expected to be around 2--3 GeV.
The presently available experimental data \cite{HPT} are not sufficient
to test this idea.
New experiments at the current electron facilities are strongly called for.
On the theoretical side, further studies on the model dependence of our
results are definitely needed.
This applies not only to the hadron quark model but also to the
microscopic description for VDM.
The latter point could be very important since we find knockout
contribution to be most distinct near the $\phi$ production threshold
while the diffractive scattering is normally associated with large $s$
and small $t$ region.
Lastly, in threshold production, the outgoing $\phi$ and proton move
with a small relative momentum and the OZI avoiding rescattering
processes deserve also to be studied.

A.I.T. is grateful to S.B. Gerasimov and S.V. Goloskokov
for useful discussions.
He would also like to thank National Science Council of ROC for support,
and the Physics Department of National Taiwan University for warm
hospitality.
Y.O. is grateful to the Alexander von Humboldt Foundation for financial
support.
This work was supported in part by NSC under grant No. NSC86-2112-M002-016.


\nopagebreak

\pagebreak

\end{document}